# A hybrid TIM-NOMA scheme for the Broadcast Channel

V. Kalokidou[1,*], O. Johnson[2] and R. Piechocki[3]

[1]Communication Systems and Networks Research Group, School of Engineering, University of Bristol, Bristol, UK.
[2]School of Mathematics, University of Bristol, Bristol, UK.
[3]Communication Systems and Networks Research Group, School of Engineering, University of Bristol, Bristol, UK.

## Abstract

Future mobile communication networks will require enhanced network efficiency and reduced system overhead. Research on Blind Interference Alignment and Topological Interference Management (TIM) has shown that optimal Degrees of Freedom can be achieved, in the absence of Channel State Information at the transmitters. Moreover, the recently emerged Non-Orthogonal Multiple Access (NOMA) scheme suggests a different multiple access approach, compared to the orthogonal methods employed in 4G, resulting in high capacity gains. Our contribution is a hybrid TIM-NOMA scheme in $K$-user cells, where users are divided into T groups. By superimposing users in the power domain, we introduce a two-stage decoding process, managing "inter-group" interference based on the TIM principles, and "intra-group" interference based on Successful Interference Cancellation, as proposed by NOMA. We show that the hybrid scheme can improve the sum rate by at least 100% compared to Time Division Multiple Access, for high SNR values.

*Corresponding author. Email: vaia.kalokidou@bristol.ac.uk.

## 1. Introduction

Future increase in the number of mobile devices, using data-hungry applications, will lead to highly dense cellular networks, demanding high capacity performance with the least possible system overhead. As mentioned in [1], in the 2020s, the volume of mobile traffic is expected to increase by a factor of 500, and future radio access should aim at a spectrum efficiency enhancement, which will be at least 3 times larger than that of Long-Term Evolution (LTE). The traditional case of Time Division Multiple Access (TDMA), since it has bounded total Degrees of Freedom (DoF), cannot constitute an optimal interference management technique for future mobile networks. Novel interference management schemes, categorised into a) interference shaping (e.g. Interference Alignment (IA), Interference Neutralization), and b) interference exploitation (e.g. Network Coding), reviewed in [2], can provide more DoF and improve the performance of communication networks.

Interference Alignment, introduced by Maddah-Ali, Motahari and Khandani in [3] and Cadambe and Jafar in [4], allows in the $K$-user interference channel $K/2$ Degrees of Freedom (DoF) to be achieved, assuming global perfect CSI. IA differs from other interference management schemes, as it attempts to align interference, rather than avoid, reduce or cancel it. However, IA requirement of full CSI is infeasible and costly. The scheme of Blind IA (BIA), presented by Wang, Gou and Jafar in [5] and Jafar in [6], for certain network scenarios, can achieve full DoF in the absence of CSI at the transmitters (CSIT), reducing considerably the system overhead. Additionally, in [7] Jafar introduces how the BIA scheme can be employed in certain cellular networks, including heterogeneous networks, by seeing frequency reuse as a simple form of IA.

Usually, when interference is strong, when compared to the desired signal, it is decoded, whereas when the interference signal is weaker than the desired link, then interference is treated as noise. In [8], Jafar introduces the Topological Interference Management (TIM) scheme, which can be considered as a form of BIA in which the position of every user in the cell(s), and therefore the strength of their channels, is taken into account. Requiring only knowledge of the network's topology at the transmitters, 1/2 DoF can be achieved for every user in the SISO Broadcast Channel (BC), by treating weak interference links as noise. Moreover, in [9] Sun and Jafar discuss the implications of increasing the number of receive antennas resulting in an increase on the network's DoF.

In [1], Saito et al. propose the Non-Orthogonal Multiple Access (NOMA) scheme for future radio access, in contrast to the Orthogonal Frequency Division Multiple Access (OFDMA) and Single Carrier-Frequency Division Multiple Access (SC-FDMA) orthogonal schemes currently adopted by 4G mobile systems. According to the NOMA scheme, multiple users are superimposed in the power domain and Successful Interference Cancellation (SIC) reception is performed at the decoding stage, ultimately improving the capacity and throughput performance. Furthermore, Benjebbour et al. in [10] present the benefits of NOMA and discuss its performance considering adaptive modulation and coding, and frequency-domain scheduling. Moreover, Ding and co-authors in [11]-[12] discuss the superior performance of NOMA in terms of ergodic sum rates and the importance of power allocation, and a cooperative NOMA scheme where users with higher channel gains have prior information about other users' messages, respectively.

In addition, Ding, Fan and Poor in [13] study user pairing on two NOMA schemes and how it affects the sum rate. The first scheme, F-NOMA, with fixed power allocation, pairs users with very distinctive channel conditions, whereas the second one, CR-NOMA, inspired by cognitive radio, pairs users with similar channel conditions. Lastly, user pairing and the performance of NOMA has been also studied from an information theory perspective, as discussed in [15], researching the relationship between the rate region achieved by NOMA and the capacity region of the BC, and showing that NOMA can outperform TDMA not only in terms of the sum rate, but for every user's rate as well, and observing, for the 2-user case, that different power allocation to users simply determines different points on the rate region graph.

In this paper, which constitutes a revised and updated version of [16], based on [1] and [8], we introduce a hybrid TIM-NOMA scheme in general $K$-user cells, extending the special case of SISO, presented in [16], to MIMO systems. Our contribution is the combination of the TIM and NOMA schemes, in a two-stage decoding way, dividing users in $T$ groups. In the first-stage, we apply the TIM scheme to manage "inter-group" interference, with no need to ignore weak interference links. In the second-stage, we employ NOMA, at every group of users separately, to manage "intra-group" interference through SIC. Finally, we discuss how the sum rate performance of the system is significantly improved with the employment of the hybrid scheme when compared to Time Division Multiple Access (TDMA).

The rest of the paper is organized as follows. Section 2 presents the general description of the hybrid scheme, with the aid of example models, including the determination of the transmit power, and the two-stage decoding process. Section 3 presents the achievable rate formula for every user in the network. Finally, Section 4 gives an overview of our results, illustrated with graphs, discussing how the users' distance from the basestation, and the amount of interference affects their performance. Section 5 summarizes the main findings of our work and discusses further developments of the hybrid scheme.

## 2. System Model

Consider the Broadcast Channel (BC) network, as shown in Figures 1 and 2, for the SISO and MIMO case respectively. At the centre of the cell, there is one transmitter $T_x$ with $N_t$ antennas, and $K$ users equipped with $N_r$ antennas each, with $N_t \geq N_r$. Transmitter $T_x$ has $L = N_r$ messages to send to every user, and moreover, when it transmits to user $k$, where $k \in \{1, 2, \ldots, K\}$, it causes interference to all the other $K - 1$ users in the macrocell. We describe an example of our scheme, where the radius of the cell is considered as $R = 5$ km and the distance of every user from the basestation is given by $d_k$.

Furthermore, users are divided into $T$ groups $\{G_1, G_2, \ldots, G_T\}$, in such a way so that, on ordering receivers by distance, there are always $T - 1$ users from the remaining $T - 1$ groups separating 2 users from the same group. This ensures we place users with considerable difference in their channel strengths in the same group. The operation is performed over $T$ time slots, over which we assume that channel coefficients remain the same. The transmitter has only knowledge of the topology of the network.

According to the NOMA scheme, described fully in [1] and [10], users are multiplexed, in the power domain, at the transmitters, and then at the receivers, signal separation is performed based on SIC. Decoding is performed based on an optimal order (in the order of decreasing channel gains divided by the power of noise and interference), resulting in every user being able to decode the signals of users coming before them in the decoding order.

The general concept of the hybrid TIM-NOMA scheme, is that every user, in order to recover its desired signal, uses the principles of TIM to manage interference coming from transmissions to users NOT belonging to their own group (i.e. their channel strengths are quite similar), and the principles of NOMA to manage interference due to transmissions to users belonging in their own group (i.e. their channel strengths are quite different).

According to our research, NOMA seems to work better when applied to users with considerable difference in their channel gains. Therefore, introducing TIM in the NOMA scheme, and splitting users into groups, provides a solution for the cases where users' gains do not differ much. The aforementioned reason, combined with fact that both schemes do not require CSIT, as discussed in [1]

# A hybrid TIM-NOMA scheme for the Broadcast Channel

and [8], results in a very smooth and successful combination of them.

## 2.1. The MIMO Broadcast Channel

In Section 2.1, we will consider an example model, depicted in Figure 1, with $K = 4$ users, $T = 2$ time slots and groups $\{G_1, G_2\}$. Users 1 and 3 are in group $G_1$, and users 2 and 4 are in group $G_2$. Finally, the users' distances from the transmitter are given by: $d_1 = 1$ km, $d_2 = 2$ km, $d_3 = 3$ km, $d_4 = 4$ km.

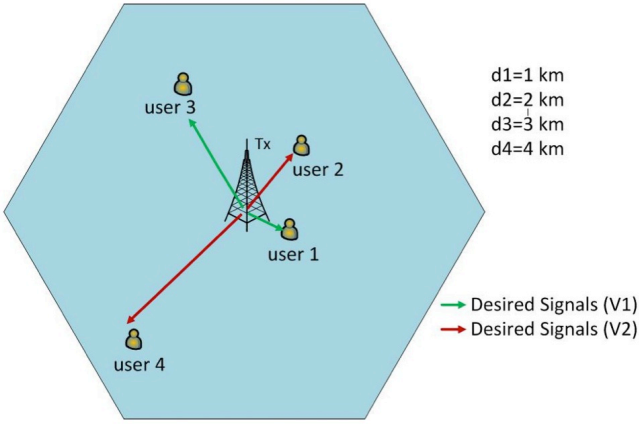

**Figure 1.** MIMO case: 4 users divided into 2 groups ($N_t = N_r = 2$). Users $\{1,3\}$ in $G_1$, users $\{2,4\}$ in $G_2$.

**Transmitted Power**

The $TN_r \times 1$ signal at receiver $k$, considering slow fading (i.e. channels is fixed through transmission time), is given by:

$$y_k = H_k x + z_k, \tag{1}$$

where $H_k \in \mathbb{C}^{TN_r \times TN_t}$ is the channel transfer matrix from $T_x$ to receiver $k$ and is given by $H_k = \sqrt{\gamma_k}(I_T \otimes h_k)$, (here and throughout $\otimes$ denotes the Kronecker (Tensor) product), with $h_k$ denoting the channel coefficients from $T_x$ to $k$ for one time slot. Due to the users' different locations, channel coefficients are statistically independent, and follow an i.i.d. Gaussian distribution $\mathcal{CN}(0,1)$. Moreover, $\gamma_k = \frac{1}{d_k^n}$ denotes the path loss, and $n$ is the path loss exponent considered for an urban environment, i.e. $n = 3$. Finally, $z_k \sim \mathcal{CN}(0, \sigma_n^2 I_{TN_r})$ denotes the independent Additive White Gaussian Noise (AWGN) at the input of receiver $k$.

Taking into consideration the position of each user $k$ in the cell, and therefore their distance $d_k$ from the basestation, ordering users increasingly, in increasing order of path loss $\gamma_k$, the following relationship (assuming all users have the same received noise power $\sigma_n^2$) follows:

$$\gamma_K > \gamma_{K-1} > \cdots > \gamma_2 > \gamma_1, \tag{2}$$

with user 1 being very close to the basestation and user $K$ at the edge of the cell. Therefore, weaker channels, of users' being far from the basestation, need to be boosted, such that the following holds for the transmit power $P_k$ of every user:

$$P_K > P_{K-1} > \cdots > P_2 > P_1. \tag{3}$$

For the case of the MIMO BC, based on our research, for the hybrid scheme, we propose two ways of allocating transmit power to each user $k$:

a) Considering the rate for every user (see Section 3) and examining the amount by which the rate deteriorates when $\sigma_n^2 \to 0$, we determine transmit power in groups and users, for the example model, as follows:

$$P_{G_1} = \left(\frac{a^2}{N_t}\right) \frac{\frac{1}{\gamma_3} + \frac{1}{\gamma_1}}{\sum_{k=1}^{4} \frac{1}{\gamma_k}} = \left(\frac{a^2}{N_t}\right) \delta_1, \tag{4}$$

$$P_{G_2} = \left(\frac{a^2}{N_t}\right) \frac{\frac{1}{\gamma_4} + \frac{1}{\gamma_2}}{\sum_{k=1}^{4} \frac{1}{\gamma_k}} = \left(\frac{a^2}{N_t}\right) \delta_2, \tag{5}$$

$$P_1 = a_1 P_{G_1}, \tag{6}$$

$$P_2 = a_2 P_{G_2}, \tag{7}$$

$$P_3 = (1 - a_1) P_{G_1}, \tag{8}$$

$$P_2 = (1 - a_2) P_{G_2}, \tag{9}$$

where $a \in \mathbb{R}$ is a constant determined by power considerations. Then, we can vary $a_1, a_2$ to find values that optimize the network's performance.

b) Considering the SINR for every user separately, and taking $\sigma_n^2 \to 0$ and then $\sigma_n^2 \to \infty$, aiming at making SINR for all users equal, we determine transmit power in groups and users, for the example model, as follows:

$$P_1 = \left(\frac{a^2}{N_t}\right) k_2 \frac{1}{\gamma_3}, \tag{10}$$

$$P_2 = \left(\frac{a^2}{N_t}\right) k_2 \frac{1}{\gamma_4}, \tag{11}$$

$$P_3 = \left(\frac{a^2}{N_t}\right) k_1 \frac{1}{\gamma_3}, \tag{12}$$

$$P_4 = \left(\frac{a^2}{N_t}\right) k_1 \frac{1}{\gamma_4}, \tag{13}$$

where

$$k_1 + k_2 = \frac{1}{\frac{1}{\gamma_3} + \frac{1}{\gamma_4}},  \quad (14)$$

so that

$$P_T = a^2. \quad (15)$$

This can be considered as a special case of a), for which:

$$a_1 = a_2 = \frac{k_2}{k_1 + k_2}, \quad (16)$$

where for matters of simplicity we define $c = \frac{k_2}{k_1+k_2}$. Again, we can vary $k_1, k_2$ to find values that optimize the network's performance.

### Stage 1: "Inter-group" interference management (TIM)

In the network, there will be $T$ precoding vectors $\boldsymbol{v}_t$, where $t \in \{1, 2, \ldots, T\}$, which are $T \times 1$ unit vectors. The choice of precoding vectors, carrying messages to users in the cell, is not unique, and we choose them in such a way so that every precoding vector $\boldsymbol{v}_t$ is orthogonal to all the remaining $T - 1$ precoding vectors.

The $T \times 1$ transmitted vector $\boldsymbol{x}$ is given by:

$$\boldsymbol{x} = \sum_{k=1}^{K} \sqrt{P_k} \left(\boldsymbol{v}_{t(k)} \otimes \boldsymbol{U}\right) \boldsymbol{x}_k, \quad (17)$$

with $t(k) \in \{1, 2, \ldots, T\}$ denoting the number of the group $G_t$ each user $k$ belongs to and $\boldsymbol{U}$ an $N_t \times L$ matrix (determining which messages will be transmitted by which antenna), with:

(i) for $N_t = N_r = L$, $\boldsymbol{U}$ is equal to the identity matrix, i.e. $\boldsymbol{U} = \boldsymbol{I}_L$,
(ii) and for $N_t > N_r, L$, $\boldsymbol{U}$ is equal to a matrix with entries 1 or 0, with the sum of the entries of its columns equal to $L$.

**Example 1.** For the example model, we choose the precoding vectors $\boldsymbol{v}_1$ and $\boldsymbol{v}_2$, for groups $G_1$ and $G_2$ respectively, as:

$$\boldsymbol{v}_1 = \begin{bmatrix} 1/2 \\ \sqrt{3}/2 \end{bmatrix}, \quad (18)$$

$$\boldsymbol{v}_2 = \begin{bmatrix} -\sqrt{3}/2 \\ 1/2 \end{bmatrix}, \quad (19)$$

where for $G_1 = \{1,3\}$ and $G_2 = \{2,4\}$, and the $2 \times 2$ matrix $\boldsymbol{U}$ is:

$$\boldsymbol{U} = \boldsymbol{I}_2, \quad (20)$$

**Theorem 1.** *Multiplying the received signal $\boldsymbol{y}_k$ with $\boldsymbol{v}_i^T \otimes \boldsymbol{S}$, where $\boldsymbol{S}$ is an $L \times N_r$ matrix that*

(i) *for $N_t = N_r = L$, $\boldsymbol{S}$ is equal to $\boldsymbol{U}$, i.e. $\boldsymbol{U} = \boldsymbol{S}$,*
(ii) *and for $N_t > N_r, L$, $\boldsymbol{S}$ is equal to a matrix with entries 1 or 0, with the sum of the entries of its columns equal to $N_r$,*

*the resulting signal at every receiver $k$, is given by:*

$$\widetilde{\boldsymbol{y}_k} = \left(\sum_{j \in G_i} \sqrt{P_j} \sqrt{\gamma_k} (\boldsymbol{S} \boldsymbol{h}_k \boldsymbol{U}) \boldsymbol{x}_j\right) + \widetilde{\boldsymbol{z}_k}, \quad (21)$$

*where $k \in G_i$, and $\widetilde{\boldsymbol{z}_k} = (\boldsymbol{v}_i^T \otimes \boldsymbol{S})\boldsymbol{z}_k$ remains white noise with the same variance.*

*Proof:* We show that $(\boldsymbol{v}_i^T \otimes \boldsymbol{S})$ removes "inter-group" interference, i.e. interference resulting from transmissions to users in groups $\{G_j\}$ for $j = 1, \ldots, T$ and $j \neq i$, at the $k$th receiver:

$$(\boldsymbol{v}_i^T \otimes \boldsymbol{S})\boldsymbol{y}_k = \left(\sum_{k=1}^{K} \sqrt{P_k} \sqrt{\gamma_k} (\boldsymbol{v}_i^T \boldsymbol{v}_{t(k)} \boldsymbol{S} \boldsymbol{h}_k \boldsymbol{U}) \boldsymbol{x}_k\right) + \widetilde{\boldsymbol{z}_k}$$

$$= \left(\sum_{j \in G_i} \sqrt{P_j} \sqrt{\gamma_k} (\boldsymbol{v}_i^T \boldsymbol{v}_i \boldsymbol{S} \boldsymbol{h}_k \boldsymbol{U}) \boldsymbol{x}_j\right) + \widetilde{\boldsymbol{z}_k}, \quad (22)$$

where by definition, for $j = 1, \ldots, T$ and $j \neq i$, $\boldsymbol{v}_i^T \boldsymbol{v}_j = 0$. ∎

**Example 2.** For the example model, for groups $G_1$ and $G_2$ respectively:

$$\boldsymbol{v}_1^T = \begin{bmatrix} 1/2 & \sqrt{3}/2 \end{bmatrix}, \quad (23)$$

$$\boldsymbol{v}_2^T = \begin{bmatrix} -\sqrt{3}/2 & 1/2 \end{bmatrix}, \quad (24)$$

The $2 \times 1$ post-processed signals at receivers are:
For $i = 1, 3$:

$$\widetilde{\boldsymbol{y}_i} = \left(\sum_{j=1,3} \sqrt{P_j} \sqrt{\gamma_i} (\boldsymbol{S} \boldsymbol{h}_i \boldsymbol{U}) \boldsymbol{x}_j\right) + \widetilde{\boldsymbol{z}_i}, \quad (25)$$

and for $i = 2, 4$:

$$\widetilde{\boldsymbol{y}_i} = \left(\sum_{j=2,4} \sqrt{P_j} \sqrt{\gamma_i} (\boldsymbol{S} \boldsymbol{h}_i \boldsymbol{U}) \boldsymbol{x}_j\right) + \widetilde{\boldsymbol{z}_i} \quad (26)$$

### Stage 2: "Intra-group" interference management (NOMA)

The concept of NOMA will be applied in each group $G_t$ separately. Based on [1, Section 3], for every group $G_t$, the SIC process is applied at every receiver. All users are ordered in increasing order of their path loss $\gamma_k$. Each user $k$ can correctly decode the signals of users, in their own group, whose path loss is larger than theirs, i.e. come before them in (2), by considering their own signal as noise. In the case where user $k$ receives interference from transmissions to users in their own group that have a smaller path loss than they do, then user $k$ simply decodes its own signal considering "intra-group" interference from users, in their own group, who come after them in (2), as noise. Maximum-Likelihood (ML) reception is performed every time a user decodes its own or another user's signal.

**Example 3.** For the example model, the decoding order for the users is:

$$\gamma_4 > \gamma_3 > \gamma_2 > \gamma_1, \tag{27}$$

In group $G_1$: Receiver 3 decodes its own signal, considering interference from transmissions to user 1 as noise.

Receiver 1 decodes first signal $x_3$ (finding $\widetilde{x_3}$), considering its own signal as noise, and subtracts the estimate $\widetilde{x_3}$ from its post-processed signal $\widetilde{y_1}$. Then, it decodes its own signal:

$$\overline{\widetilde{y_1}} = \widetilde{y_1} - (\boldsymbol{v}_1^T \otimes \boldsymbol{S})\sqrt{\gamma_1}(I_T \otimes \boldsymbol{h}_1)(\boldsymbol{v}_1 \otimes \boldsymbol{U})\widetilde{x_3}. \tag{28}$$

If $\widetilde{x_3} = x_3$, then (28) reduces to the interference-free channel.

In group $G_2$: Receiver 4 decodes its own signal, considering interference from transmissions to user 2 as noise.

Finally, receiver 2 decodes signal $x_4$ (finding $\widetilde{x_4}$), considering its own signal as noise, and substracts the estimate $\widetilde{x_4}$ from its post-processed signal $\widetilde{y_2}$. Then, it decodes its own signal:

$$\overline{\widetilde{y_2}} = \widetilde{y_2} - \boldsymbol{v}_2^T \sqrt{\gamma_2}(I_T \otimes \boldsymbol{h}_2)\boldsymbol{v}_2 \widetilde{x_4}. \tag{29}$$

## 2.2. Special Case: The SISO Broadcast Channel

The special case of the SISO BC was introduced in [16], considering the example model, depicted in Figure 2, with $K = 5$ users, $T = 2$ time slots and groups $\{G_1, G_2\}$. Users 1, 3 and 5 are in group $G_1$, and users 2 and 4 are in group $G_2$. Finally, the users' distances from the transmitter are given by: $d_1 = 0.5$ km, $d_2 = 1.5$ km, $d_3 = 2.5$ km, $d_4 = 3.5$ km, $d_5 = 4.5$ km.

**Transmitted Power**
Taking into consideration the position of each user $k$ in the cell, and therefore its distance $d_k$ from the basestation, and assuming that all users have the same received noise power $\sigma_n^2$ (2) holds, and therefore, as (3) shows, weaker channels, of users' being far from the basestation, need to be boosted.

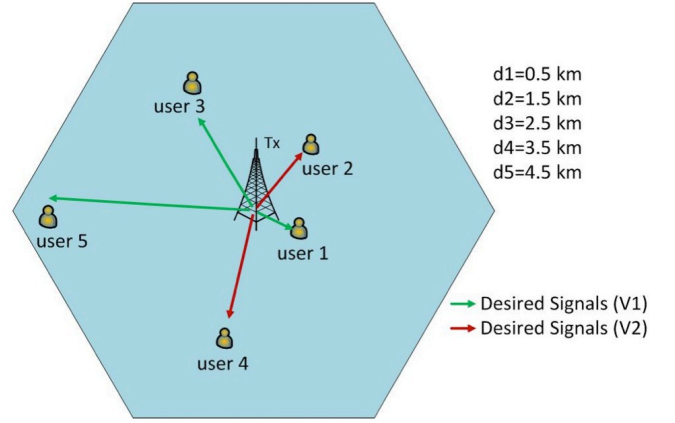

**Figure 2.** SISO case: 5 users divided into 2 groups. Users $\{1,3,5\}$ in $G_1$ and users $\{2,4\}$ in $G_2$.

The energy of the input symbol vector $x_k \in \mathbb{C}$, of each user $k$, is defined as:

$$\mathbb{E}[|x_k|^2] = 1. \tag{30}$$

For every user $k$ in the cell, we choose to take its transmitted power given by:

$$P_k = a^2 \frac{d_k^2}{\sum_{j=1}^K d_j^2}, \tag{31}$$

where $a \in \mathbb{R}$ is a constant determined by power considerations. The total transmit power is given by the power constraint:

$$P_T = \sum_{j=1}^K P_j\, norm(x_k) = a^2. \tag{32}$$

**Stage 1: "Inter-group" interference management (TIM)**
For the SISO case, the $T \times 1$ transmitted vector $\boldsymbol{x}$ is given by:

$$\boldsymbol{x} = \sum_{k=1}^K \sqrt{P_k}\, \boldsymbol{v}_{t(k)} x_k, \tag{33}$$

with $t(k) \in \{1, 2, \dots, T\}$ denoting the number of the group $G_t$ each user $k$ belongs to, and $\boldsymbol{v}_{t(k)}$ determined as described in Section 2.1.

**Example 4.** For the example model, the precoding vectors $\boldsymbol{v}_1$ and $\boldsymbol{v}_2$, for groups $G_1$ and $G_2$ respectively, are given by (18) and (19), and the $2 \times 1$ transmitted vector is:

$$\boldsymbol{x} = \sum_{k=1}^5 \sqrt{P_k}\, \boldsymbol{v}_{t(k)} x_k,$$

$$\text{(34)}$$

where for $G_1 = \{1,3,5\}$ and $G_1 = \{2,4\}$.

**Theorem 2.** *Multiplying the received signal $\boldsymbol{y}_k$ with the transpose of the precoding vector $\boldsymbol{v}_i$, the resulting signal at every receiver k, is given by:*

$$\widetilde{\boldsymbol{y}_k} = \boldsymbol{v}_i^T \boldsymbol{H}_k \left( \sum_{j \in G_i} \sqrt{P_j}\, \boldsymbol{v}_i x_j \right) + \widetilde{\boldsymbol{z}_k}$$

$$= \sqrt{\gamma_k}\, h_k \left( \sum_{j \in G_i} \sqrt{P_j}\, x_j \right) + \widetilde{\boldsymbol{z}_k},$$

$$\text{(35)}$$

where $k \in G_i$, and $\widetilde{\boldsymbol{z}_k} = \boldsymbol{v}_i^T \boldsymbol{z}_k$ remains white noise with the same variance.

*Proof:* We show that $\boldsymbol{v}_i^T$ removes "inter-group" interference, i.e. interference resulting from transmissions to users in groups $\{G_j\}$ for $j = 1, \dots, T$ and $j \neq i$, at the $k$th receiver:

$$\boldsymbol{v}_i^T \boldsymbol{y}_k = \boldsymbol{v}_i^T \left( \sqrt{\gamma_k}(I_T \otimes h_k) \sum_{k=1}^{K} \sqrt{P_k}\, \boldsymbol{v}_{t(k)} x_k + \boldsymbol{z}_k \right)$$

$$= \boldsymbol{v}_i^T \sqrt{\gamma_k}(I_T \otimes h_k) \left( \sum_{j \in G_i} \sqrt{P_j}\, \boldsymbol{v}_i x_j \right) + \boldsymbol{v}_i^T \boldsymbol{z}_k,$$

$$\text{(36)}$$

where by definition, for $j = 1, \dots, T$ and $j \neq i$, $\boldsymbol{v}_i^T \boldsymbol{v}_j = 0$. ∎

**Example 5.** For the example model, the 1×1 post-processed signals at receivers are:
For $i = 1, 3, 5$:

$$\widetilde{\boldsymbol{y}_i} = \boldsymbol{v}_1^T \boldsymbol{H}_i \left( \sum_{j=1,3,5} \sqrt{P_j}\, \boldsymbol{v}_1 x_j \right) + \boldsymbol{v}_1^T \boldsymbol{z}_i,$$

$$\text{(37)}$$

and for $i = 2, 4$:

$$\widetilde{\boldsymbol{y}_i} = \boldsymbol{v}_2^T \boldsymbol{H}_i \left( \sum_{j=2,4} \sqrt{P_j}\, \boldsymbol{v}_2 x_j \right) + \boldsymbol{v}_2^T \boldsymbol{z}_i$$

$$\text{(38)}$$

**Stage 2: "Intra-group" interference management (NOMA)**

The concept of NOMA is applied in each group $G_t$ separately, following the same procedure as described in Section 2.1.

**Example 6.** For the example model, the decoding order for the users is:

$$\gamma_5 > \gamma_4 > \gamma_3 > \gamma_2 > \gamma_1,$$

$$\text{(39)}$$

In group $G_1$: Receiver 5 decodes its own signal, considering interference from transmissions to users 1 and 3 as noise.

Receiver 3 decodes first signal $x_5$ (finding $\widetilde{x_5}$), considering its own signal as noise, and subtracts the estimate $\widetilde{x_5}$ from its post-processed signal $\widetilde{\boldsymbol{y}_3}$. Then, it decodes its own signal:

$$\overline{\widetilde{\boldsymbol{y}_3}} = \widetilde{\boldsymbol{y}_3} - \boldsymbol{v}_1^T \sqrt{\gamma_3}(I_T \otimes h_3) \boldsymbol{v}_1 \widetilde{x_5}.$$

$$\text{(40)}$$

Receiver 1 decodes first signal $x_5$ (finding $\widetilde{x_5}$) and then $x_3$ (finding $\widetilde{x_3}$), subtracting every time the estimate of the interfering signal from its post-processed one, considering its own signal as noise, eventually decoding its own, interference-free, signal:

$$\overline{\widetilde{\boldsymbol{y}_1}} = \left( \widetilde{\boldsymbol{y}_1} - \boldsymbol{v}_1^T \sqrt{\gamma_1}(I_T \otimes h_1) \boldsymbol{v}_1 \widetilde{x_5} \right)$$
$$- \boldsymbol{v}_1^T \sqrt{\gamma_1}(I_T \otimes h_1) \boldsymbol{v}_1 \widetilde{x_3}.$$

$$\text{(41)}$$

In group $G_2$: Receiver 4 decodes its own signal, considering interference from transmissions to user 2 as noise.

Finally, receiver 2 decodes first signal $x_4$ (finding $\widetilde{x_4}$), considering its own signal as noise, and substracts the estimate $\widetilde{x_4}$ from its post-processed signal $\widetilde{\boldsymbol{y}_2}$. Then, it decodes its own signal:

$$\overline{\widetilde{\boldsymbol{y}_2}} = \widetilde{\boldsymbol{y}_2} - \boldsymbol{v}_2^T \sqrt{\gamma_2}(I_T \otimes h_2) \boldsymbol{v}_2 \widetilde{x_4}.$$

$$\text{(42)}$$

## 3. Achievable Rate

For the MIMO BC, since there is no CSIT, the total rate for each user $k$, in group $G_t$, per time slot, setting $\boldsymbol{M}_k = \boldsymbol{S} h_k \boldsymbol{U}$, and $\rho_{t(k)} = \sum_{\substack{j \in G_t \\ j < k}} \left\| \sqrt{\gamma_k}\, h_k \boldsymbol{v}_t \right\|^2 P_j$, is given by:

$$R_k = \frac{1}{T} \mathbb{E}\left[ \log \det \left( \boldsymbol{I}_{N_r} + \frac{P_k}{N_t(\rho_{t(k)} + \sigma_n^2)} \gamma_k \boldsymbol{M}_k \boldsymbol{M}_k^T \right) \right],$$

$$\text{(43)}$$

where $k \in G_t$.

If only one user is active (TDMA), with all other users shut down, the achievable rate, per time slot, is given by:

$$R_k = \frac{1}{T} \mathbb{E}\left[ \log \det \left( \boldsymbol{I}_{N_r} + \frac{P_T}{N_t \sigma_n^2} \gamma_k \boldsymbol{h}_k \boldsymbol{h}_k^T \right) \right].$$

$$\text{(44)}$$

For the special case of the SISO BC, since there is no CSIT, the total rate for each user $k$, in group $G_t$, per time slot, setting $D = \sum_{k=1}^{K} d_k^2$, is given by:

$$R_k = \frac{1}{T} \log \left( 1 + \frac{P_T}{\sum_{\substack{j \in G_t \\ j < k}} |\boldsymbol{H}_k \boldsymbol{v}_t|^2 P_j + \sigma_n^2} \frac{d_k^2}{D} |\boldsymbol{v}_t^T \boldsymbol{H}_k \boldsymbol{v}_t|^2 \right),$$

$$\text{(45)}$$

**A hybrid TIM-NOMA scheme for the Broadcast Channel**

where $k \in G_t$.

If only one user is active (TDMA), with all other users shut down, the achievable rate, per time slot, is given by:

$$R_k = \frac{1}{T}\log\left(1 + \frac{P_T}{\sigma_n^2}|\boldsymbol{H}_k\boldsymbol{v}_t|^2\right), \quad (46)$$

**Example 7.** For the example MIMO model the achievable rate, for every user, since $\boldsymbol{M}_k = \boldsymbol{h}_k$, is given by:

$$R_1 = \frac{1}{2}\log\det\left(\boldsymbol{I}_2 + \frac{P_T}{N_t\sigma_n^2}a_1\delta_1\gamma_1\boldsymbol{h}_1\boldsymbol{h}_1^T\right), \quad (47)$$

$$R_2 = \frac{1}{2}\log\det\left(\boldsymbol{I}_2 + \frac{P_T}{N_t\sigma_n^2}a_2\delta_2\gamma_2\boldsymbol{h}_2\boldsymbol{h}_2^T\right), \quad (48)$$

$$R_3 = \frac{1}{2}\log\det\left(\boldsymbol{I}_2 + \frac{P_T}{N_t(\rho_1 + \sigma_n^2)}(1-a_1)\delta_1\gamma_3\boldsymbol{h}_3\boldsymbol{h}_3^T\right), \quad (49)$$

$$R_4 = \frac{1}{2}\log\det\left(\boldsymbol{I}_2 + \frac{P_T}{N_t(\rho_2 + \sigma_n^2)}(1-a_2)\delta_2\gamma_4\boldsymbol{h}_4\boldsymbol{h}_4^T\right), \quad (50)$$

where $\rho_1 = \left\|\sqrt{\gamma_3}\boldsymbol{h}_3\boldsymbol{v}_1\right\|^2 P_1$ and $\rho_2 = \left\|\sqrt{\gamma_4}\boldsymbol{h}_4\boldsymbol{v}_2\right\|^2 P_2$.

Similar expressions hold in the SISO case, and are given by (23)-(27) in [16].

## 4. Overview of Results

Our simulations were based on the example models already described and were performed in Matlab. The statistical model chosen was i.i.d. Rayleigh and our input symbols were Quadrature Phase Shift Keying (QPSK) modulated. Maximum-Likelihood (ML) detection was performed in the end of the decoding stage. The total transmit power was considered as $40W$ (a typical value for transmit power in macrocells for 4G systems), and therefore $a$, a constant determined by power considerations in (15) and (32), is given by $a = \sqrt{40}$. Moreover, simulations were performed for 100-750 frames, with each frame consisting of 6144 bits.

### 4.1. Degrees of Freedom

For the SISO case, in [9], with the TIM scheme, the DoF that can be achieved for every user are 0.5 DoF, i.e. one message sent over two time slots. In [1], with the NOMA scheme, 1 DoF can be achieved for every user.

Introducing the hybrid scheme, the total DoF in the network are given by:

$$DoF_{total} = K\frac{N_r}{T}, \quad (51)$$

where $\frac{K}{T}$ is the average number of users per group, showing that the less the number of groups is, the more DoF are provided.

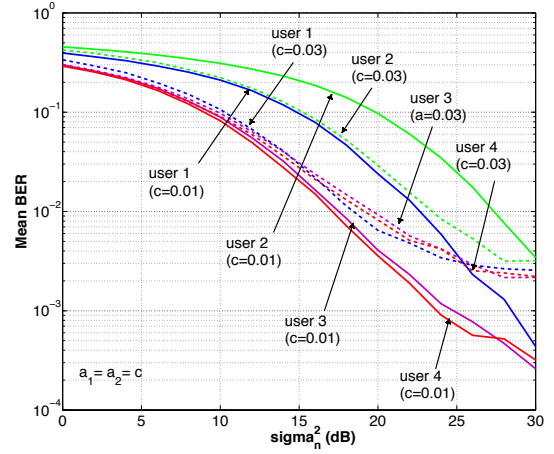

**Figure 3.** BER Performance of every user in the network (MIMO BC) for $c = \{0.01, 0.03\}$. For $c = 0.03$ BER of each users tends to be the same for high SNR values.

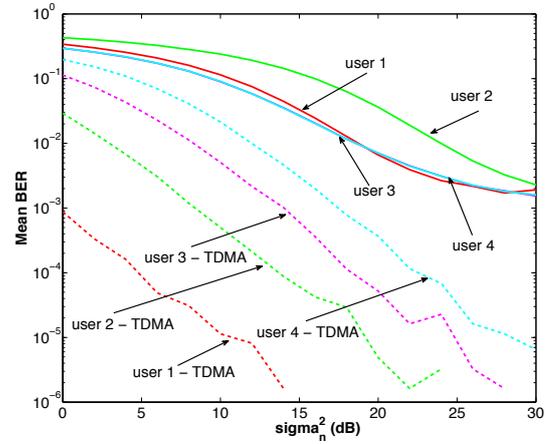

**Figure 4.** BER Performance of every user compared to TDMA (MIMO BC) for $c = 0.0255$. BER performance is better for the case that only one user is active (TDMA).

### 4.2. Bit Error Rate (BER) Performance

First of all, the BER performance of our example model was investigated. Based on our findings, the distance of every user $k$ from the transmitter is a key feature that determines the BER performance of every user. Furthermore, the amount of transmit power allocated to each user has also a considerable impact on the BER performance.

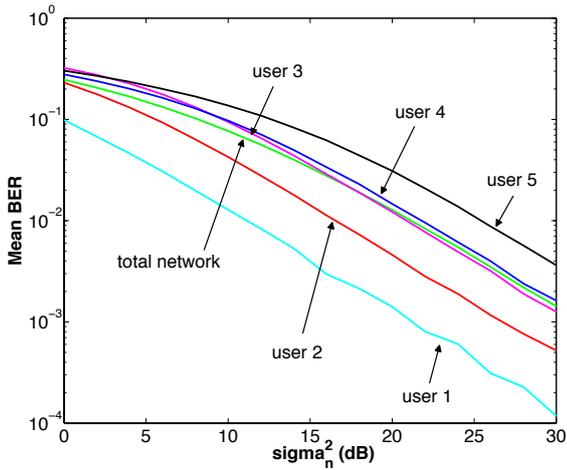

**Figure 5.** BER Performance of the total network and every user separately (SISO BC). The closer a user is to the basestaion, the better their BER performance.

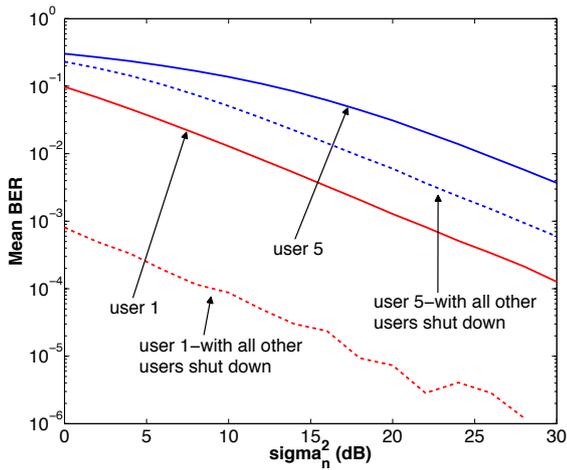

**Figure 6.** BER Performance of users compared to TDMA (SISO BC). BER performance is better for the case that only one user is active (TDMA).

For the MIMO channel, the special case of $a_1 = a_2 = \frac{k_2}{k_1+k_2} = c$ was considered. In Figure 3, it can be observed that for different values of $c$, the BER performance of the network changes, proving its dependence on the way power is allocated. For $c = 0.03$, the performance of each user is very similar. Moreover, Figure 4 shows a comparison between the BER performance in the case of the hybrid scheme and the TDMA one. Generally, BER performances are better in the case of TDMA, however the hybrid schemes, for $c = 0.0255$, provides a fairer performance to network users, as their BER tends to be similar for high SNR values.

For the SISO channel, where we can consider a fixed power allocation scheme, as depicted in Figure 5, users who are closer to the basestation, like users 1 and 2, have a better performance than users who are far from the basestation, like users 4 and 5. Furthermore, in Figure 6, in which for matters of simplicity only users 1 and 5 are studied, as the performances of the remaining users lie in between, it can be observed that BER performances are better when only one user is active (TDMA). Furthermore, the closer a user is to the transmitter, the less improvement we observe in their performance, in the case where all other users are inactive.

### 4.3. Rate Performance

The rate of the network will be a function of the user's distance from the basestation and the amount of interference considered as noise, if any, as shown in (43) and (45). In Figures 7 and 8, for $c = 0.0255$, it can be observed that the rate decreases with the distance of the user from the transmitter and the amount of interference considered as noise. In particular, user 1, who is the closest to the basestation and manages all interference during the decoding stage, achieves the best rate performance. On the contrary, user 4 (MIMO case) and user 5 (SISO case), who are the furthest from the basestation and consider all "intra-group" interference as noise, achieve the worst performance overall for high SNR values.

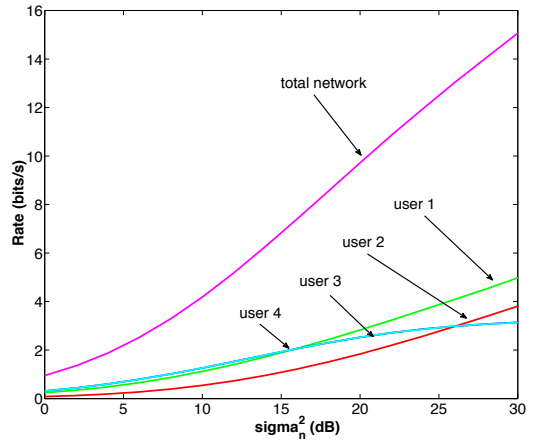

**Figure 7.** Rate Performance of the network and every user separately (MIMO BC) for $c = 0.0255$. For high SNR values, the closer a user is to the basestation, the better their rate performance is.

The rate performance of the hybrid scheme was compared to the rate users would achieve if only one was active (TDMA), as given by (44) and (46) for the MIMO and SISO cases respectively. In general, as it can be observed in Figures 9 and 10, rate performances are better

**A hybrid TIM-NOMA scheme for the Broadcast Channel**

when only one user is active. However, it can be observed that for high SNR values the sum rate of the hybrid scheme outperforms the TDMA average rate.

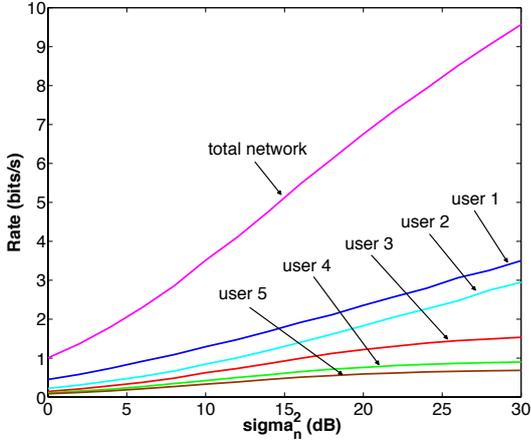

**Figure 8.** Rate Performance of the network and every user separately (SISO BC). The closer a user is to the basestation, the better their rate performance is.

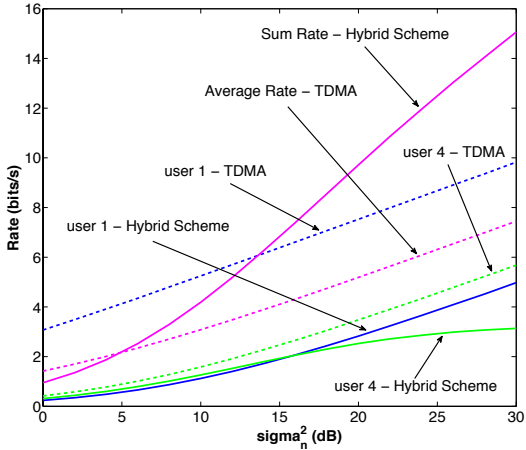

**Figure 9.** Rate Performance compared to TDMA (MIMO BC) for $c = 0.0255$. For SNR values >5dB, the hybrid schemes outperforms TDMA.

Finally, in order to emphasize the gain, in terms of sum rate, the hybrid scheme provides, this gain is depicted in Figures 11 and 12 for the cases of MIMO and SISO respectively, where the value of the ratio

$$R = \frac{R_{hybrid}}{R_{TDMA}},$$
(52)

where $R_{hybrid}$ is the sum rate of the hybrid scheme and $R_{TDMA}$ the sum rate of TDMA, is studied for a range of SNR values. In the MIMO case, it can be observed that the hybrid scheme outperforms TDMA by more than 100% for values of SNR greater than 26dB. In the SISO case, for values greater than 11dB the hybrid scheme achieves at least double the rate than would be achieved by TDMA.

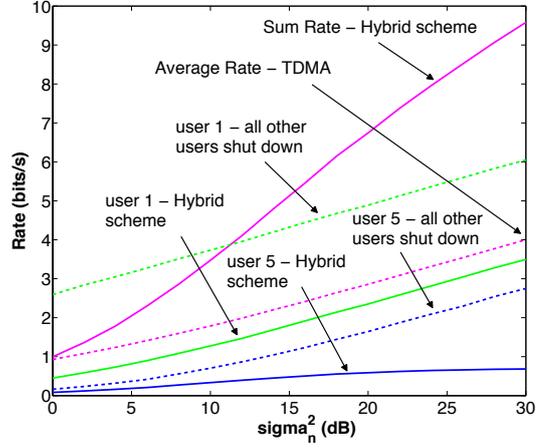

**Figure 10.** Rate Performance compared to TDMA (SISO BC). For SNR values >0dB, the hybrid schemes outperforms TDMA.

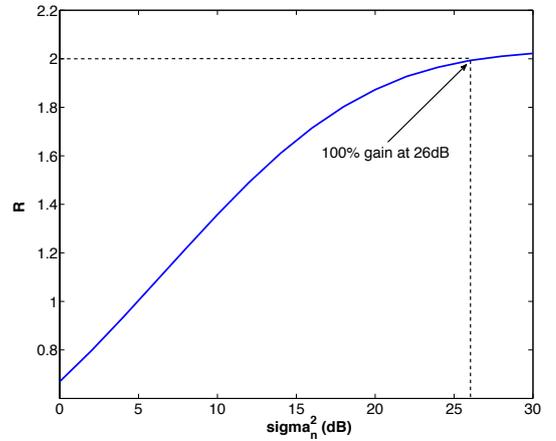

**Figure 11.** Ratio of sum rate of hybrid scheme over sum rate of TDMA (MIMO BC). For SNR values >26dB, the employment of the hybrid scheme provides 100% gain over TDMA.

## 5. Summary

Overall, this paper introduces a novel hybrid scheme that can be employed in the MIMO BC of a cell, with $K$ users divided into $T$ groups. The hybrid scheme combines basic principles of the TIM and NOMA schemes, by treating "inter-group" interference and "intra-group" interference separately and by a different method. Moreover, the employment of TIM in the cases where users' gains do not differ much, solves performance issues that were faced by NOMA, which performs better when the channel gain difference among users is large. Furthermore, the system's complexity is reduced, providing flexibility, when compared to the NOMA scheme, without decreasing the rate performance that the system would have if NOMA was only applied. In general, the employment of the proposed scheme results in high data rates, very good BER performance, and reduced system overhead (due to the absence of CSIT requirement). Most interestingly, for SNR values greater than 26dB for the MIMO case and 11dB for the SISO case, the total sum rate of the hybrid scheme is 100% better than the sum rate of TDMA, proving the gain in terms of sum rate the hybrid scheme results in.

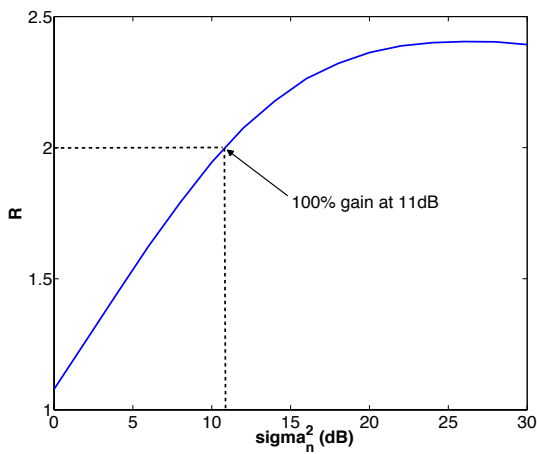

**Figure 12.** Ratio of sum rate of hybrid scheme over sum rate of TDMA (SISO BC). For SNR values >11dB, the employment of the hybrid scheme provides 100% gain over TDMA.

The simple concept of the hybrid TIM-NOMA scheme introduced in this paper, as an extension of our work presented in [16], suggests that it could be employed in dense networks, and potentially in heterogeneous networks once certain adjustments in the algorithm are made. Finally, apart from the special case of the SISO channel where a fixed power allocation scheme can be employed, the determination of a fair power allocation method, as mentioned in [14], constitutes a complex procedure in NOMA schemes and requires exhaustive simulations that will maximize the system's performance. However, as stated in [15] and based on our simulations different power allocation gives different points on the rate region and therefore the power allocation scheme can vary according to the performance requirements of every system the hybrid scheme is employed in.

**Acknowledgements.**
This work was supported by NEC; the Engineering and Physical Sciences Research Council [EP/I028153/1]; and the University of Bristol. The authors thank Simon Fletcher & Patricia Wells for useful discussions.